\documentclass[conference]{IEEEtran}
\IEEEoverridecommandlockouts

\usepackage{graphicx}
   \graphicspath{{./Figures/}}
\usepackage{hyperref}
\usepackage{threeparttable}
\usepackage{tabularx}
\usepackage{amsmath}    
\usepackage{algorithm}
\usepackage{algorithmic}
\usepackage{multirow}
\usepackage{cite}
\usepackage{svg}
\usepackage{color}
\usepackage{adjustbox}
\usepackage{multirow,tabularx}
\usepackage{makecell}
\DeclareGraphicsExtensions{.pdf,.jpeg,.png,.fig, .emf}
   \usepackage{subfigure}
    \usepackage{epstopdf}
    \usepackage{epsfig}
    \usepackage{subfigure}
    \usepackage{epstopdf}
    \usepackage{epsfig}
\usepackage{xspace}
\newcommand{\design}{TAP-CAM\xspace}    
 \usepackage{booktabs}

\begin{document}
\bstctlcite{IEEEexample:BSTcontrol}

% \title{Energy-Efficient Mismatch Detection in 64-Column Arrays Based on Ferroelectric Ternary Content Addressable Memories for 0-6 Bits}
\title{TAP-CAM: A Tunable Approximate Matching Engine based on Ferroelectric Content Addressable Memory}

\author{ 
\small
Chenyu Ni$^1$, Sijie Chen$^1$, Che-Kai Liu$^2$, Liu Liu$^3$, Mohsen Imani$^4$, Thomas Kämpfe$^5$, Kai Ni$^3$, \\
Michael Niemier$^3$, Xiaobo Sharon Hu$^3$,  Cheng Zhuo$^{1,6,*}$, Xunzhao Yin$^{1,6,*}$\\
$^1$Zhejiang University, Hangzhou, China; 
$^2$Georgia Institute of Technology, GA, USA\\
$^3$University of Notre Dame, IN, USA;
$^4$University of California Irvine, CA, USA\\
$^5$Fraunhofer IPMS, Dresden, Germany\\
$^6$Key Laboratory of Collaborative Sensing and Autonomous Unmanned Systems of Zhejiang Province, China\\
$^*$Corresponding authors, email: \{czhuo, xzyin1\}@zju.edu.cn

    }

\maketitle
%\pagestyle{empty}
%\vspace{-3mm}

% \fi
% \IEEEpeerreviewmaketitle

\begin{abstract}

Pattern search is crucial in numerous analytic applications for retrieving data entries akin to the query. Content Addressable Memories (CAMs), an in-memory computing fabric, directly compare input queries with stored entries through embedded comparison logic, facilitating fast parallel pattern search in memory.
While conventional CAM designs offer exact match functionality, they are inadequate for meeting the approximate search needs of emerging data-intensive applications. 
Some recent CAM designs propose approximate matching functions, but they face limitations such as excessively large cell area or the inability to precisely control the degree of approximation. 
In this paper, we propose TAP-CAM, a novel ferroelectric field effect transistor (FeFET) based ternary CAM (TCAM) capable of both exact and tunable approximate matching. 
TAP-CAM employs a compact 2FeFET-2R cell structure as the entry storage unit, %for basic storage and computing functions at the unit circuit level, enabling a dense CAM array to enhance energy efficiency. 
and similarities in Hamming distances between input queries and stored entries are measured using an evaluation transistor associated with the matchline of CAM array. 
%Extensive Monte Carlo simulations assess the impact of FeFET device variation. 
The operation, robustness and performance of the proposed design at array level have been discussed and evaluated, respectively. 
We conduct a case study of K-nearest neighbor (KNN) search to benchmark the proposed TAP-CAM at application level.
%Results demonstrate that TAP-CAM achieves a 6.78× energy improvement compared to 2FeFET CAM implementing approximate match functionality, and 16.95× compared to 16T CMOS CAM implementing exact match functionality and 2FeFET CAM implementing approximate match functionality, along with a 3.06\% accuracy enhancement. 
Results demonstrate that compared to 16T CMOS CAM with exact match functionality, TAP-CAM achieves a 16.95$\times$ energy improvement, along with a 3.06\% accuracy enhancement. Compared to 2FeFET TCAM with approximate match functionality, TAP-CAM achieves a 6.78$\times$ energy improvement.

\end{abstract}

%\vspace{-3ex}
%New addition: the latency is reduced to <3ns and the EDP/cell is reduced to 0.286EDP(fJ)/bit(under 64X64), harvesting 333X and 90.5X (both conservative) improvement for latency and EDP/cell respectively(comp. Nature comm.).

%\vspace{-1em}

\section{Introduction}
\label{sec:introduction}

In the era of advancing artificial intelligence, the computational demands on AI models are rapidly increasing. 
Training data volumes across various domains like computer vision (CV) \cite{CV}, natural language processing (NLP) \cite{neural}, and speech recognition \cite{speechrecognition} have surged, posing significant challenges to computing hardware and architectures, both at the edge and in data centers. The traditional von Neumann architecture, with its constant data movement between memory and processing units, exacerbates energy consumption and latency issues, intensifying the ``Memory Wall" problem.
To tackle this challenge, emerging computing paradigms, notably In-Memory Computing (IMC), have gained attention. 
IMC directly employs parallel data operations within the memory, enhancing core performance and efficiency while alleviating the ``Memory Wall" problem \cite{IMC2,  IMC3, yin2024deep, yin2024ferroelectric, yang2024energy, li2024febim}. 

Content Addressable Memory (CAM) emerges as a hardware solution of IMC, enabling parallel and efficient searching and similarity  measurement  within the memory. 
CAMs compare input data with all stored data simultaneously, and output the stored entry that matches with input or has the highest similarity to the input.
%avoiding power consumption and latency associated with data transfers. Therefore, CAMs 
Therefore, CAMs are viewed as a potential solution for accelerating 
%the processor-memory bottleneck. 
%Traditionally used in network routers and associative processors, CAMs are now gaining interest in 
various data-centric workloads like bioinformatics \cite{zhong2023asmcap,laguna2020seed}, machine learning \cite{2FeFETa, xu2024ferex, hu2021memory}, and neural language processing \cite{neural}.
%As a specialized solution to the Memory Wall problem, CAMs utilize the entire memory array to accelerate parallel search operations, showing great potential in today’s computing networks. 
Specifically, CAMs significantly speed up Hyperdimensional Computing (HDC), making this brain-inspired computing paradigm  efficient for tasks like image classification and speech recognition \cite{HDC1, HDC2, HDC3}. 
This effectiveness arises from CAMs' ability to transform sequential pattern matching into highly parallelizable computational tasks and simplify the complex distance measurements into Hamming distance \cite{kim2020geniehd}. 
The rapid search and matching capability of CAMs make them essential components in applications requiring efficient data access and retrieval.

Conventional  CMOS based CAM design consists of 10-16 transistors per cell, which results in large area overhead and high energy consumption \cite{16TCMOS}. 
%These problems severely limits their applicability \cite{16TCMOS} in various applications. 
To tackle the area and energy challenges,
%faced by CMOS CAM designs, 
researchers have proposed utilizing emerging non-volatile memory (NVM) devices to construct more compact and efficient CAM designs, as these  CAMs merge the storage and logic within the NVM devices, thus offering significant area and energy saving. 
%aiming to improve the  performance, area and energy efficiency. 
CAMs based on 2-terminal NVMs like resistive RAM (RRAM) \cite{li2021sapiens,Chang3t1r}, magnetic tunneling junction (MTJ) \cite{Matsunaga4t2mtj, zhuo2022design}, phase change memory (PCM) \cite{jing2t2r}, and 3-terminal ferroelectric field effect transistor (FeFET) \cite{2FeFET,4T2FeFET,1FeFET1R-transfer, yin2022ferroelectric, yin2023ultracompact, xu2023challenges, Huang2024, yin2020fecam, li2020scalable} have been explored. 
Among these devices, FeFETs stand out in constructing the compact and efficient CAM designs due to their unique hysteresis I-V characteristics, high current ON/OFF ratio, high off-state resistance, low write energy, and compatibility with CMOS technology \cite{Liu2022eva-cam}.  
While non-volatile storage can achieve high area efficiency and  mitigate the high energy consumption caused by CMOS technology, these CAMs still encounter limitations for data-intensive applications  due to their exact search functionality.
In the era of big data, as the amount of data for processing bursts and the chances of exact matching drop down,  these CAMs with limited array size fail to maintain the hardware utilization efficiency while consuming extra area and energy overheads.
%the application of approximate matching is increasingly common, offering potential hardware utilization efficiency improvements while maintaining an acceptable level of accuracy.
Many applications require approximate pattern search functions where entries with a similarity within a certain threshold distance to the search query are desired. 
To address the challenge of limited CAM utilization efficiency, 
various CAM designs implementing approximate pattern search have been proposed. 
These approximate CAMs improve the utilization and overall energy efficiency by compensating the search accuracy within an acceptable range. 
For instance, HD-CAM \cite{conventionalCAM} introduced a 10T CMOS-based approximate CAM with a matchline (\textit{ML}) charge redistribution technique, but it suffers from a large cell area and lacks the support for wildcard (\textit{don’t care}) bits.
Moreover, the  design is unable to precisely control the degree of approximation, bit-by-bit.
MHCAM \cite{liu2023reconfigurable} presented an approximate CAM design based on FeFET with programmable thresholds, but it's tailored to applications requiring multi-state Hamming distance. 
\cite{MASC} implemented threshold matching by leveraging voltage scaling and controlling the precharge period, but its high energy consumption and inability to precisely control the threshold limit its applications.
\cite{2FeFETa} introduced approximate matching capabilities using 2FeFET TCAM. It computes the Hamming distance between search and stored vectors in a highly parallelized manner by monitoring \textit{ML} discharge rate. Despite achieving notable energy efficiency and density in TCAM, it lacks fine-grained control over approximate search precision.
To address aforementioned challenges of existing approximate CAMs, in this work, we propose TAP-CAM, a general approximate matching engine featuring a bit-by-bit tunable threshold match function. 
We consider FeFET as a representative NVM device, and propose to utilize a novel 2FeFET-2R ternary CAM (TCAM) cell structure to store ternary value. 
An evaluation transistor is employed between the parallel connected TCAM cells and the CAM array sense amplifier to control the \textit{ML} discharge rate, and the tunable threshold of the approximate matching functionality is set by the bias voltage of the evaluation transistor. 
We validate the bit-wise XNOR logic and the tunable  threshold matching functionality of  TAP-CAM design at cell and array levels,  respectively,  and conduct extensive Monte Carlo simulations to examine the robustness against device-to-device variations.
We use the K-nearest neighbor search (KNN) as a representative application to investigate the benefits of TAP-CAM at application level.
Evaluation results demonstrate that TAP-CAM achieves a 16.95$\times$ energy improvement and 3.06\% accuracy improvement compared to 16T CMOS CAM with exact match function. Compared to 2FeFET TCAM with
approximate match functionality, TAP-CAM achieves a 6.78$\times$
energy improvement.

The rest of paper is organized as follows: Sec.~\ref{sec:background} reviews the FeFET device characteristics and existing CAM designs. Sec.~\ref{sec:proposed_work} introduces the proposed TAP-CAM. Sec.~\ref{sec:eval} presents the evaluation results and the KNN case study. Finally, Sec.~\ref{sec:conclusion} summarizes the paper.

\section{Background}
\label{sec:background}

In this section, we discuss the structure and operational principles of FeFETs, and review existing CAM design works.

\begin{figure}%[H]
    \centering
    \includegraphics[width=1\linewidth]{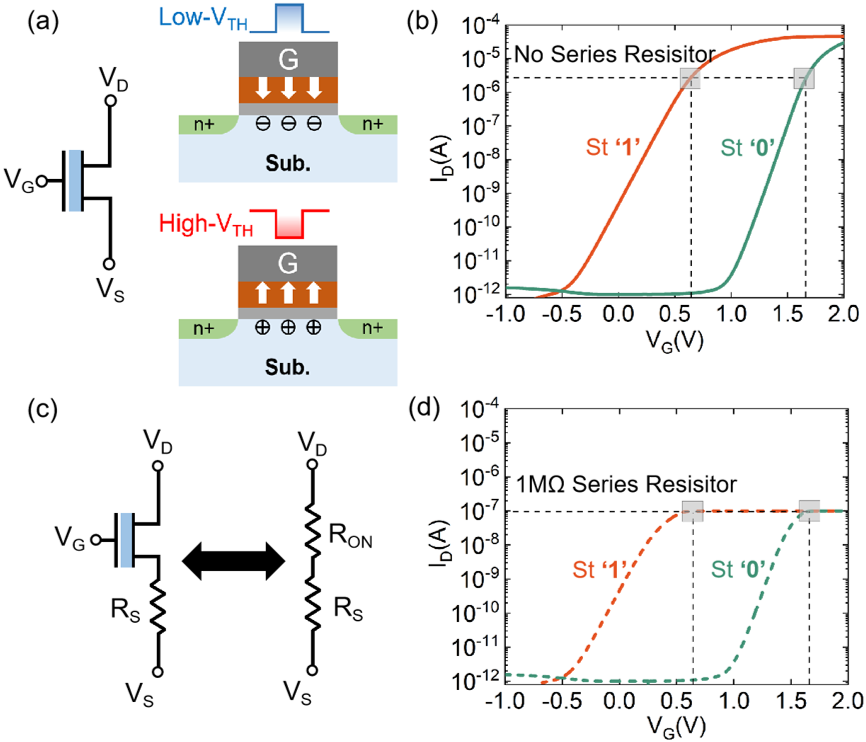}
   % \vspace{-0.4cm}
    \caption{\textbf{(a)} FeFET polarization directions and channel conditions after memory write operations;  \textbf{(b)} The FeFET $\textit{I}_\textit{D}$-$\textit{V}_\textit{G}$ characteristics after positive/negative gate write; % Source is grounded; 
    \textbf{(c)} 1FeFET-1R structure and equivalent circuit; \textbf{(d)} The 1FeFET-1R $\textit{I}_\textit{D}$-$\textit{V}_\textit{G}$ characteristics after positive/negative gate write.
    %Source is grounded.
    }

    \label{fig:fefet}
   %  \vspace{-0.4cm}
\end{figure}

%\vspace{-2ex}
%\vspace{-0cm}
\subsection{FeFET Basics}
\label{sec:device}
\setlength{\abovecaptionskip}{2pt}
\setlength{\belowcaptionskip}{2pt}

Recent advancements in ferroelectric material, particularly hafnium oxide ($\text{HfO}_\text{2}$), have spurred research interest in ferroelectric transistors  and the development of non-volatile circuit designs compatible with CMOS technology \cite{yin2020fecam}. 
FeFETs 
%belong to the subclass of metal-oxide-semiconductor field-effect transistors (MOSFETs) and 
incorporate a ferroelectric 
(FE) layer  within the gate stack. These devices exhibit unique electrical hysteresis characteristics, exhibiting reversible polarization states upon an applied voltage-driven electric field. 
%Integration of a ferroelectric capacitor with the MOSFET gate capacitor confers FeFETs with adjustable hysteresis characteristics. 
The FE layer induces a shift in the threshold voltage of the FeFET depending on the orientation of FE polarization \cite{FeFET-capacitor}, enabling non-volatile (NV) storage capabilities. 
By applying gate voltage pulses, such as -4V/+4V, to a FeFET device, as depicted in \autoref{fig:fefet}(a), it can be programmed to store low and high $\textit{V}_\textit{TH}$ states corresponding to logic ‘0’ and ‘1’, respectively. 
The associated hysteresis  $\textit{I}_\textit{D}$-$\textit{V}_\textit{G}$ transfer characteristics are shown in \autoref{fig:fefet}(b) \cite{transfer-characteristics}. FeFETs, being voltage-driven for read and write operations, exhibit superior energy efficiency compared to two-terminal current-driven NVMs.

When the FeFET operates as a current source, its ON current gradually increases with the rise in gate voltage, as depicted in \autoref{fig:fefet}(b). Consequently, there's a certain variability in the conduction current regarding the gate read voltage. 
To ensure stable ON current during operation and enhance the design robustness, a current limiter is connected to the source of the FeFET, as shown in the equivalent circuit of \autoref{fig:fefet}(c).
Prior studies \cite{1FeFET1R-transfer, yin2023ultracompact} have shown that a series resistor on the drain/source of a FeFET can regulate the ON current, with 1FeFET-1R integration experimentally demonstrated \cite{area}. Such integration suppresses the ON current variability, making it independent of the $\textit{V}_\textit{TH}$ state and gate voltage when the series resistor is sufficiently large. 
The transfer characteristic curve of the 1FeFET-1R structure is depicted in \autoref{fig:fefet}(d).  
We adopt the 1FeFET-1R structure using a series resistor as a current limiter in this work. 
This approach mitigates the impact of ON current variability on \textit{ML} discharging in a CAM array 
%and reduces high energy consumption, 
achieving low power consumption and robust tunable approximate matching functionality.

\subsection{Existing CAM Designs}
\label{sec:existing_work}

\begin{figure}
    \centering
    \includegraphics[width=\linewidth]{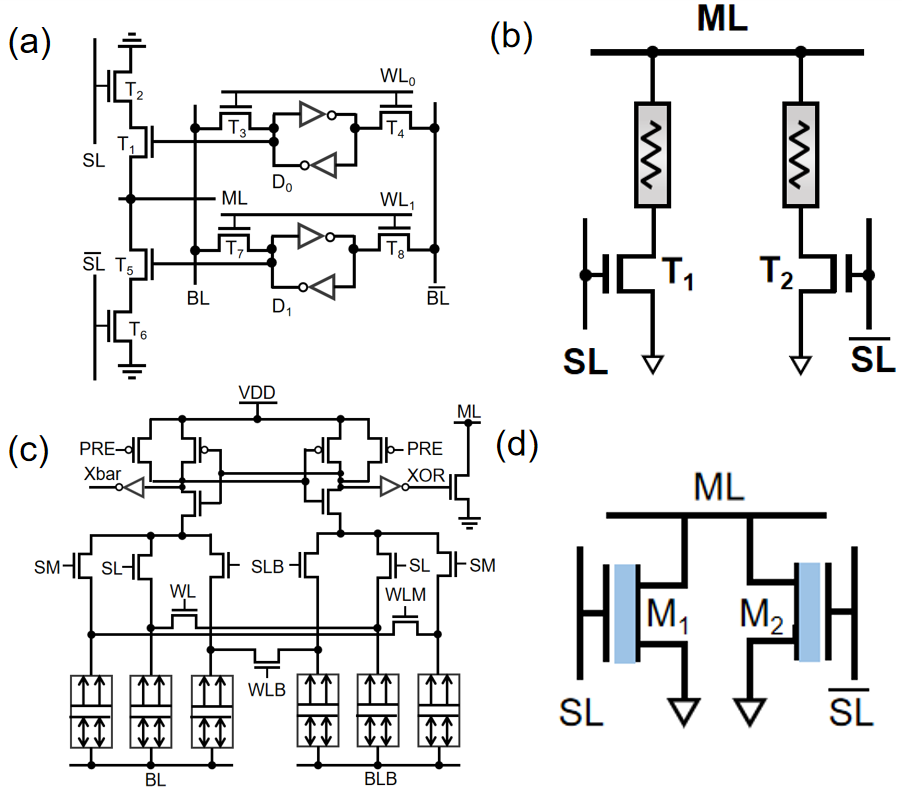}
  %  \vspace{-0.4cm}
    \caption{Schematics of \textbf{(a)} 16T CMOS TCAM cell; \textbf{(b)} 2T-2ReRAM TCAM cell; \textbf{(c)} 20T-6MTJ TCAM cell; \textbf{(d)} 2FeFET TCAM cell.}
  %  \vspace{-0.4cm}
\label{fig:CAM}
\end{figure}

%\begin{figure}
%    \centering
%    \includegraphics[width=\linewidth]{Figures/conventional_CAM.pdf}
%    \caption{(a)Architecture of an M × N TCAM array.(b)NOR-type CAM bitcell.}
%\label{fig:array}
%\end{figure}

Various CAM designs have been proposed based on CMOS technology and NVM devices. A conventional 16T CMOS TCAM cell is shown in \autoref{fig:CAM}(a). CAMs leveraging NVM typically demonstrate enhanced performance over CMOS-based counterparts. For example, a 2T-2R TCAM design based on ReRAM was proposed in \cite{jing2t2r} for its compact structure, as shown in \autoref{fig:CAM}(b). While it consumes less area compared with conventional CMOS-based CAM designs, %issues arise primarily due to 
the low HRS/LRS ratio, low variable resistance and current-driven write-in mechanism associated with large access transistors  make the write and search energy significant concerns. 
\cite{20T6MTJ} proposed a 20T-6MTJ TCAM design as illustrated in \autoref{fig:CAM}(c), greatly enhancing the search speed and search performance. However, the reduced sense margin caused by the limited TMR ratio of STT-MRAM necessitates numerous transistors to address this issue, thus severely impacting area and power consumption.

Among NVM based CAM designs, utilizing FeFET stands out due to its high ON/OFF current ratio, efficient voltage-driven write mechanisms, low energy consumption, and cost-effectiveness, enabling significant performance improvements compared to conventional CMOS designs and other NVM-based designs. Building upon advanced FeFET models, researchers have proposed various FeFET CAM designs, particularly designs of TCAM. 
The 2FeFET TCAM design as depicted in \autoref{fig:CAM}(d) offers a compact alternative than CMOS counterparts \cite{2FeFET}. 2FeFET TCAM features a smaller cell area, reduced write and search energy consumption, and search delay. 
However, it faces limitations such as the lack of support for approximate matching functionality. 

\subsection{Threshold Matching Concepts and Related Works}
\label{sec:existing_work}

\begin{figure}
    \centering
    \includegraphics[width=\linewidth]{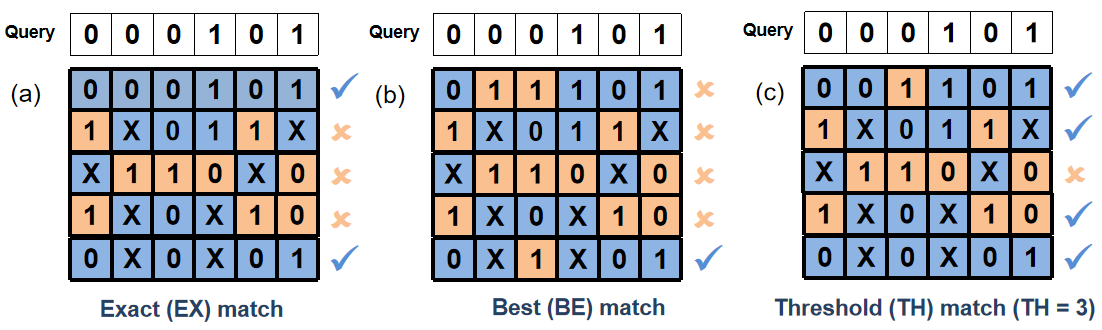}
  %  \vspace{-0.4cm}
    \caption{\textbf{(a) Exact match:} The stored entry that matches exactly with the query; \textbf{(b) Best match:} The stored entry that has the smallest distance to the query; \textbf{(c) Threshold match:} The stored entry whose distance to the query is below specified thresholds.}
 %   \vspace{-0.4cm}
\label{fig:Matchstyle}
\end{figure}

Most CMOS and NVM based CAM designs discussed earlier prioritize exact matching, as depicted in \autoref{fig:Matchstyle}(a), 
limiting their adaptability for data-intensive applications. 
%For data-centric applications,
In contrast, approximate matching gains favor due to its potential to enhance hardware utilization while maintaining acceptable accuracy. 
As a means to achieve approximate matching, best match CAMs, as illustrated in \autoref{fig:Matchstyle}(b), aim to output the stored entry with the highest similarity to the search query. 
For example, A-HAM \cite{AHAM} evaluates similarities across stored entries and identifies the closest Hamming distance to the input query. 
4T-2MTJ utilizing STT-MRAM \cite{STTMRAM} measures similarity between input query and stored entries in terms of \textit{ML} current and outputs the entry with the highest similarity. \cite{bestmatch} introduced a CAM design for minimum Hamming distance search using digital circuits for bit comparison. A Winner-Take-All (WTA) circuit at the output selects the entry with the highest degree of matching to the search query. However, CAMs designed for best matching may fail in applications requiring the output of multiple entries with specific similarities. Therefore, threshold matching CAMs were devised. 

Threshold matching CAMs, as illustrated in \autoref{fig:Matchstyle}(c), aim to provide multiple stored entries with similarity within a predefined Hamming distance (HD) threshold. 
For instance, the HD-CAM proposed in \cite{conventionalCAM} utilizes a 10T CMOS-based design incorporating \textit{ML} charge redistribution, enabling threshold matching with large HD tolerance, notably used in virus DNA classification. However, the SRAM based HD-CAM cell incurs substantial area and energy overheads. Furthermore, its effectiveness is limited in discerning patterns with substantial HDs due to the intricate tuning of \textit{ML} discharge current, making bit-by-bit tuning of HD thresholds impractical.
\cite{liu2023reconfigurable} introduced MHCAM, a multi-state CAM design encoding multiple CAM cells into distinct multi-states per dimension to perform both dimension-wise exact matching and reconfigurable threshold matching. However, additional transistors introduce fixed bit precisions (1-bit/2-bit/4-bit/8-bit per dimension), restricting fine-grained tunability in threshold matching and adaptability to applications demanding multi-state HD. The ReRAM-based CAM proposed in \cite{MASC} implements threshold matching by leveraging voltage scaling and controlling the precharge period. However, the current-driven mechanisms of ReRAMs result in high power consumption during operation and limited HD thresholds can be achieved due to the large \textit{ML} discharge current and non-trivial threshold-associated period sampling. \cite{2FeFETa} implements approximate matching functionality based on 2FeFET TCAM. It calculates the HD between search and stored vectors in a parallel manner by sensing the discharge rate of \textit{ML}. While achieving high energy efficiency and density in TCAM, it lacks precise control over the degree of approximate searching.

These threshold search CAMs all face a common issue, that they cannot precisely control the degree of approximate matching. Therefore, our design will focus on implementing bit-by-bit tuning of threshold to control the degree of approximate matching.

\section{Proposed TAP-CAM Design}
\label{sec:proposed_work}
In this section, we present the TAP-CAM design with bit-by-bit tunable HD threshold match functionality, exploiting the 2FeFET-2R structure and incorporating a threshold-defined evaluation transistor. 
We first discuss the structure and operation principles of the cell, %particularly focusing on the 2FeFET-2R configuration. Subsequently, we
and then elucidate the threshold approximate match implementation at the array level.

\subsection{2FeFET-2R TCAM Cell} 

\begin{figure}
    \centering
    \includegraphics[width=\linewidth]{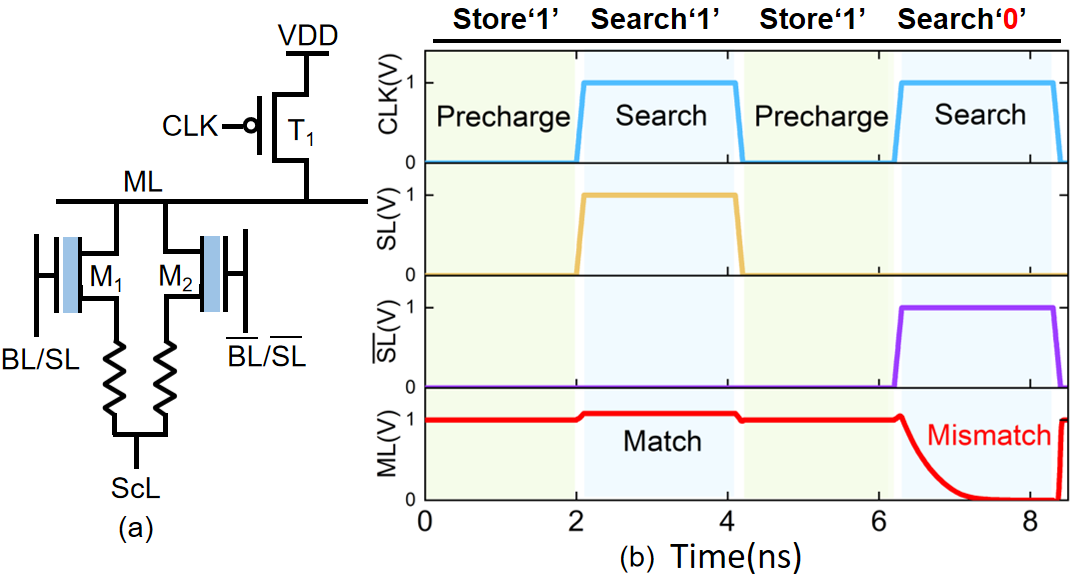}
    %\vspace{-0.4cm}
    \caption{\textbf{(a)} Structure of the proposed 2FeFET-2R TCAM cell; \textbf{(b)} Transient voltage waveforms of 2FeFET-2R CAM cell storing `1’.}
   % \vspace{-0.4cm}
\label{fig:2FeFET-2R Cell}

\end{figure}

\begin{table}[!t]
    \centering
    \caption{OPERATIONS OF 2FEFET-2R TCAM CELL}
\begin{adjustbox}{center}
\resizebox{1\columnwidth}{!}{
            \begin{tabular}{|c c |c|c|c|c|c|}
            
              \hline \hline
              $\textit{V}_\textit{write}$ = 4V & $\textit{V}_\textit{search}$ = 1V & \textit{BL}/$\overline{\textit{SL}}$ &  $\overline{\textit{BL}}$/$\textit{SL}$ & \textit{ScL} & $\textit{M}_\text{1}$ & $\textit{M}_\text{2}$   \\ 
              \hline
    \multirow{2}*{Write`1'} & Step1     &$\textit{V}_\textit{write}$ & 0 & 0 &`1' & hold\\ & Step2     &$\textit{V}_\textit{write}$ & 0 & $\textit{V}_\textit{write}$ & hold & `0'\\ 
    \hline
    \multirow{2}*{Write`0'} & Step1    & 0 & $\textit{V}_\textit{write}$ & $\textit{V}_\textit{write}$ &`0'  & hold\\
                               & Step2    &0 & $\textit{V}_\textit{write}$ & 0 & hold & `1'\\
                       \hline
              \multicolumn{2}{|c|}{Write \textit{don't care}}  &$\textit{V}_\textit{write}$ & $\textit{V}_\textit{write}$ & 0 & `1' & `1'\\
              \hline
              \hline
            \end{tabular}
             
      }
        \end{adjustbox}
    
    \label{tab:write opetarion}
\end{table}

\autoref{fig:2FeFET-2R Cell}(a) shows the structure of the proposed 2FeFET-2R TCAM Cell. It comprises a pair of parallel  1FeFET-1R structures, with the FeFET drain connected to the matchline (\textit{ML}), and the other end of the structure connected to the sourceline (\textit{ScL}), driven by either $V_{\textit{write}}$ or \textit{GND}.
The FeFET gate connects to the bitline and searchline (\textit{BL}/\textit{SL} and $\overline{\textit{BL}}$/$\overline{\textit{SL}}$). 
By adjusting the write gate input, the FeFET threshold aligns with different storage values. 
%$T_{1}$ precharges $ML$ before searching. 
The 2FeFET-2R structure can store logic `1', `0', and 
\textit{don't care} wildcard state.
%\autoref{fig:2FeFET-2R Cell}(a) shows the structure of the proposed 2FeFET-2R TCAM Cell. The 2FeFET-2R unit circuit comprises a pair of parallel 1FeFET-1R structures, where the drain of the FeFET is connected to the matchline ($ML$), and the other end of the resistor connected to the sourceline ($ScL$) can be driven to $V_{write}$ or $GND$. The gate of the FeFET is connected to the bitline and searchline ($BL$/$\overline{SL}$ and $\overline{BL}$/$SL$). By varying the input to the gate, the threshold of the FeFET can be regulated to correspond to different storage values. $T_{1}$ is connected to $ML$ for precharging $ML$ before the search. The 2FeFET-2R structure can store logic `1' and `0', and $don't\text{ $care$ } $cases. 
\autoref{tab:write opetarion} outlines the write operations of the 2FeFET-2R cell. 
Data bits are written in two steps, storing complementary logic states in each FeFET. 
To write logic `1', $V_{\textit{write}}$ is applied to \textit{BL}/\textit{SL}, while `0' to \textit{ScL} and $\overline{\textit{BL}}$/$\overline{\textit{SL}}$. This sets $V_{\textit{GS}}$ of $\textit{M}_\text{1}$ to 4V, writing logic `1' to $\textit{M}_\text{1}$. In the second step, $V_{\textit{write}}$ is applied to \textit{ScL}, while gate voltage remains the same, writing logic `0' to $\textit{M}_\text{2}$. Thus, the complementary stored  values represents logic `1'.
Similarly, to write logic `0' into the cell, `0' is written to $\textit{M}_\text{1}$ and `1' to $\textit{M}_\text{2}$, respectively. 
To write \textit{don't care} state, logic `1' is written to both $\textit{M}_\text{1}$ and $\textit{M}_\text{2}$. This sets both FeFETs to high-$\textit{V}_\textit{TH}$ state, matching regardless of the search value, aligning with the masking function of `\textit{don't care}' bits.
During writes, \textit{ML} is grounded to eliminate static current. \autoref{fig:fefet}(b) displays $\textit{I}_\textit{D}$-$\textit{V}_\textit{G}$ curves for FeFETs under different write pulses.

%\autoref{tab:write opetarion} summarizes the write operations of the 2FeFET-2R unit circuit. The data bits are written into the two FeFETs in the TCAM cell in two steps, storing opposite logic values in each FeFET. In order to write logic `1', in the first step, $V_{write}$ is applied to $BL$/$\overline{SL}$, while `0' is applied to $ScL$ and $\overline{BL}$/$SL$. Therefore, the gate-source voltage ($V_{GS}$) of $M_1$ is 4V, switching the FE polarization within FeFET and writing logic `1' to $M_1$. In the second step, $V_{write}$ is applied to $ScL$, while the gate voltage of the two FeFETs remains the same as in the first step (i.e. 4V at $BL$/$\overline{SL}$ and 0 at $\overline{BL}$/$SL$). As a result, the $V_{GS}$ of $M_2$ is -4V, and the logic `0' is written to $M_2$. Thus, the 2FeFET-2R unit as a whole represents the storage of logic '1'. 

%Similarly, to write logic `0' into the TCAM cell, we write `0' to $M_1$ and `1' to $M_2$, respectively. To write $don't\text{ $care$ }$into the TCAM cell, it only takes one step to write logic `0' to both $M_1$ and $M_2$. In this way, both FeFETs corresponding to `don't care' are in low-$V_{TH}$ state, resulting in a match regardless of the search value, which aligns with the masking function of `don't care' bits. During writing operations, the matchline $ML$ needs to be driven to ground to eliminate the influence of static current on the $ML$ voltage. \autoref{fig:fefet}(b) displays the $I_{D}$-$V_{G}$ curves corresponding to different write pulses of FeFETs.

During search, \textit{ML} voltage is precharged to high via a precharge transistor, and the search voltages are applied to searchlines ($\textit{SL}$/$\overline{\textit{SL}}$) according to the query data. For logic `1', \textit{SL} set to 1V, and 0 for logic `0', the  \textit{ML} voltage indicates the matching result. 
\autoref{fig:2FeFET-2R Cell}(b) validates the function of the 2FeFET-2R cell. 
\textit{ML} is first precharged by controlling $\textit{T}_\text{1}$'s gate voltage \textit{CLK}, and then left floating upon  search phase. 
When searching `1', \textit{ML} voltage stays high with  \textit{SL} = 1V, indicating a match. Conversely, searching `0' rapidly drops \textit{ML} voltage to 0, indicating a mismatch. %Results in \autoref{fig:2FeFET-2R Cell}(b) align with expectations, validating unit circuit's storage and computing functions.

%During the search operation, the voltage of $ML$ is previously precharged to a high level through a precharge transistor, and different search voltages are applied to the searchlines ($SL$/$\overline{SL}$) according to the input data. Here we set $SL$ to 1V for logic `1' and 0 for logic `0', and observe the voltage change of $ML$ to reflect the matching result. As shown in \autoref{fig:2FeFET-2R Cell}(b), by controlling the gate voltage $CLK$ of $T_{1}$ to precharge $ML$, the precharging is halted after entering the search phase. When searching for logic '1', the voltage on $ML$ remains almost unchanged at a high level when $SL$ voltage is 1V, indicating a match between the search query and the stored entry. Conversely, when searching for logic '0', the voltage on $ML$ rapidly drops to 0, indicating a mismatch between the search query and the stored entry. The operation results shown in \autoref{fig:2FeFET-2R Cell}(b) are consistent with expectations, validating the correctness of the unit circuit's storage and computing functions.

\subsection{2FeFET-2R TCAM Array}

\begin{figure}
    \centering    
    \includegraphics[width=\linewidth]{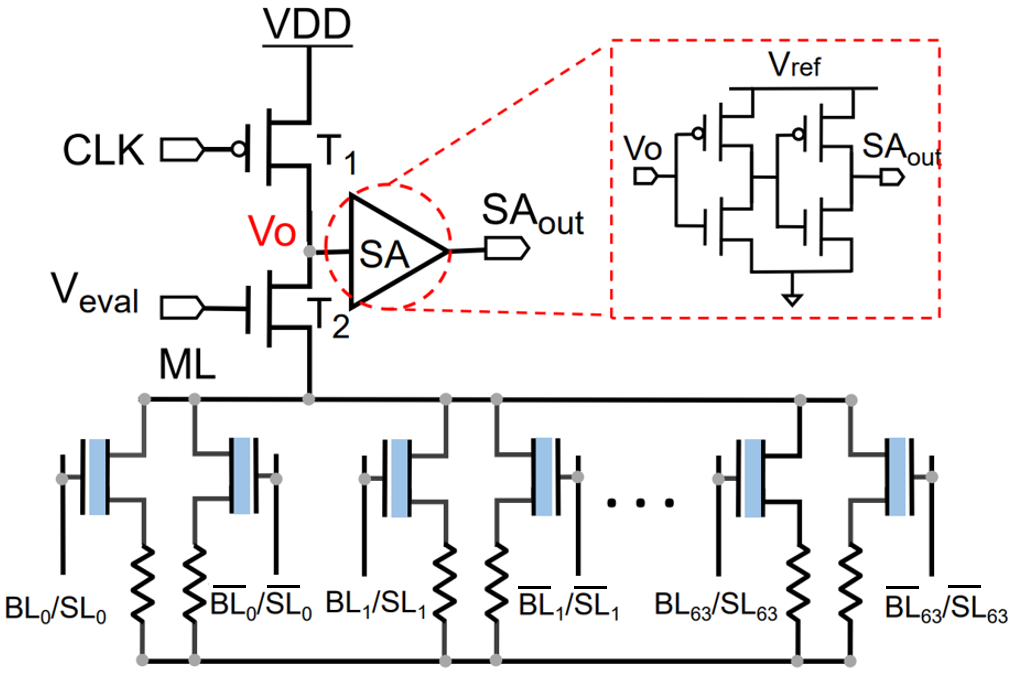}
    \caption{Structure of a 2FeFET-2R TCAM array with wordlength 64.}
\label{fig:1x64}
%\vspace{-0.4cm}
\end{figure}

\autoref{fig:1x64} demonstrates the schematic of the proposed 2FeFET-2R TAP-CAM array storing a 64-bit word with corresponding peripheral circuits. 
%This array stores a 64-bit word, sharing $ML$ and $ScL$ across all 64 cells.
PMOS $\textit{T}_\text{1}$ precharges \textit{ML} before the search operation, while an evaluation transistor $\textit{T}_\text{2}$ is connected between \textit{ML} and $\textit{V}_\text{o}$ to enable tunable threshold matching function. 
%$T_2$ evaluates the $ML$ voltage for threshold match function. 
Adjusting the gate voltage of the evaluation transistor controls the discharge rate of \textit{ML}, allowing varying mismatch bits to be sensed by the sense amplifier (SA) as a match case. 
%A sense amplifier (SA) detects the  output, enhancing signal stability, forming output waveform. 

%to mitigate $ML$ parasitic capacitance impact on sensing time.
%As shown in \autoref{fig:1x64}, the 2FeFET-2R unit circuit is expanded into a 1×64 array and accompanied by corresponding peripheral circuits. This array has the capability to store a 64-bit word, with all 64 units sharing the same $ML$ and $ScL$. The $ML$ is connected to two transisitors, where PMOS $T_1$ is used to precharge $ML$ before the matching operation begins, while NMOS $T_2$ is connected to ML as a transistor for voltage evaluation, enabling threshold-controlled approximate match functionality. By adjusting the gate voltage applied to NMOS, the rate of $ML$ voltage decrease can be controlled, thereby achieving control over the permissible number of mismatched unit bits. Additionally, a sense amplifier(SA) is serially connected at the output end to shape and amplify the output results, thereby enhancing the stability of the output signal, ultimately forming the output waveform. During the precharging phase, the PMOS control signal $CLK$ is driven to a low level, while $V_{eval}$ is kept at a high level to ensure the conduction of both $T_1$ and $T_2$. This allows $ML$ to be precharged to $V_{DD}$ to mitigate the impact of parasitic capacitance of $ML$ on the sensing time. 

\label{sec:2FeFET-2R TCAM Array}

During the precharge, \textit{CLK} is set to low, turning  $\textit{T}_\text{1}$ and $\textit{T}_\text{2}$ ON, and precharging \textit{ML} to \textit{VDD}.
During the search phase, setting the \textit{CLK} signal high turns $\textit{T}_\text{1}$ OFF and cutting the charging path. 
Pre-defined bias voltages are applied to the gate of evaluation transistor  $\textit{V}_\textit{eval}$ based on required mismatch thresholds. 
A mismatch between the stored entry and the search query forms a conduction path from $\textit{V}_\text{o}$ to \textit{GND}, discharging $\textit{V}_\text{o}$ and decreasing the voltage.
The rate of voltage decrease depends on the number of mismatched cells and $\textit{T}_\text{2}$'s gate voltage $\textit{V}_\textit{eval}$. 
This rate affect  the output of SA $\textit{SA}_\textit{out}$ which indicates the time for $\textit{SA}_\textit{out}$ to transition from high to low. 
With constant $\textit{V}_\textit{eval}$, more mismatched bits increase the discharge current from $\textit{V}_\text{o}$ to \textit{GND}, accelerating $\textit{SA}_\textit{out}$ voltage drop. 
Similarly, with constant mismatched bits, higher $\textit{V}_\textit{eval}$ boosts the conduction of $\textit{T}_\text{2}$, hastening $\textit{SA}_\textit{out}$ voltage drop. Hence, given the fixed SA sense time, decreasing the $\textit{V}_\textit{eval}$ allows for increasing the mismatch threshold.
Without loss of generality, 
for the TAP-CAM with n bits mismatch threshold (Th-n), i.e., $\leq$n mismatch bits are sensed as a match case, and $\ge$(n+1) bits mismatch indicates a mismatch, the sense margin between the n bits mismatch and (n+1) bits mismatch is determined by the equivalent resistance and associated \textit{ML} capacitance of the array $\textit{C}_\textit{M}$.
%, as illustrated in \autoref{fig:fefet}(c), $R_{ON}$ signifies FeFET's equivalent conduction resistance, while $C_M$ represents the circuit's equivalent capacitance.
%As previously articulated, the 1FeFET-1R configuration can limit the magnitude of the conduction current and enhance 
%the robustness of the circuit. Moreover, as the resistance value of the series current limiter increases, the current gradually decreases, leading to improved stability of the circuit. In addition, we also have observed that the presence of $R_S$ results in a more significant voltage drop of $ML$ due to greater mismatched bit count. Taking n bits mismatch and (n+1) bits mismatch as examples, as illustrated in \autoref{fig:fefet}(c), $R_{ON}$ represents the equivalent conduction resistance of the FeFET, while $C_M$ represents the equivalent capacitance of the circuit. 
The equivalent resistance for the two mismatch cases can be expressed as follows:
%We can express the relationship as follows:
\begin{equation}
\label{eq:R-C circuit1}
      \textit{R}_\textit{n} = \frac{\text{1}}{\textit{n}}\cdot (\textit{R}_\textit{ON}+\textit{R}_\textit{S})
\end{equation}
\begin{equation}
\label{eq:R-C circuit2}
   %   R_{n+1} = \frac{1}{n+1}\cdot (R_{ON}+R_S)
     \textit{R}_{\textit{n}\text{+1}} = \frac{\text{1}}{\textit{n}\text{ + 1}}\cdot (\textit{R}_\textit{ON}+\textit{R}_\textit{S})
\end{equation}
where $\textit{R}_\textit{n}$ represents the approximate equivalent resistance of array with n bits mismatch, and $\textit{R}_\text{n+1}$ represents the approximate equivalent resistance of array with (n+1) bits mismatch. %$C_M$ is the associated ML capacitance.
$\textit{R}_\textit{ON}$ represents the equivalent resistance of an ON FeFET, and $\textit{R}_\textit{S}$ represents the series resistance. 
From charging and discharging formula of RC circuit, we can approximately formulate the \textit{ML} voltage \textit{U}:
\begin{equation}
\label{eq:R-C circuit3}
      \textit{U}=\textit{U}_\text{0}\cdot \textit{e}^{-\frac{\textit{t}}{\textit{RC}_\textit{M}}}
\end{equation}

\begin{equation}
\label{eq:R-C circuit4}
      \frac{\textit{dU}}{\textit{dt}}=\textit{U}_\text{0}\cdot (-\frac{\text{1}}{\textit{RC}_\textit{M}})\textit{e}^{-\frac{\textit{t}}{\textit{RC}_\textit{M}}}
\end{equation}
where $\textit{U}_\text{0}$ represents the initial voltage of \textit{ML}. 
From \autoref{eq:R-C circuit4} we can conclude that the rate of \textit{ML} voltage drop will be faster as the equivalent resistance decreases. 
%Due to the fact that the parallel resistance of n identical resistors is greater than the parallel resistance of n+1 identical resistors,
From \autoref{eq:R-C circuit1} and \autoref{eq:R-C circuit2}, $\textit{R}_\textit{n}$ is larger than $\textit{R}_{\textit{n}\text{+1}}$. 
Therefore,  the voltage of \textit{ML} corresponding to (n+1) bits mismatch drops faster than that of n bits mismatch.
Upon the sensing, the sense margin of Th-n $\Delta U$ can be expressed as follows:
%$U_n$ represents the $ML$ voltage corresponding to  n  bits mismatch, and $U_{n+1}$ represents the $ML$ voltage corresponding to the circuit with (n+1) mismatched bits. As indicated by  \autoref{eq:R-C circuit3}, we can express this relationship as follows:
\begin{equation}
\label{eq:R-C circuit5}
     \Delta \textit{U}=\textit{U}_\textit{n}-\textit{U}_{\textit{n}\text{+1}}=\textit{U}_\text{0}\cdot (\textit{e}^{-\frac{\textit{t}}{\textit{R}_\textit{n}\textit{C}_\textit{M}}}-e^{-\frac{t}{\textit{R}_{\textit{n}\text{+1}}\textit{C}_\textit{M}}}) 
\vspace{0.1cm}
\end{equation}
where $\textit{U}_\textit{n}$ represents the \textit{ML} voltage corresponding to  n bits mismatch, and $\textit{U}_{\textit{n}\text{+1}}$ represents the \textit{ML} voltage corresponding to (n+1) bits mismatch.
%With an increase in $R_S$, $\Delta$$U$ also broadens, resulting in a wider sensing margin.
From \autoref{eq:R-C circuit5}, we observe that $\textit{R}_\textit{S}$ affects the magnitude of $\Delta$$U$ over time t, thus influencing the sense margin. Simultaneously, a larger $\textit{R}_\textit{S}$ value introduces larger search delay. Therefore, selecting an appropriate $\textit{R}_\textit{S}$ value is necessary to ensure that both sense margin and search delay remain within reasonable limits. 
We here select $\textit{R}_\textit{S}$ = 0.3M.% and \textit{VDD} = 0.6V

%As the series resistance $R_S$ increases, $\Delta$$U$ also experiences an augmentation, thereby resulting in a broader sensing margin. However, this enhancement in circuit performance is accompanied by an increase in latency. Therefore, it is necessary to carefully balance both performance and latency when determining the resistance value of $R_S$.Taking all factors into account, we ultimately opt for R=0.3M and $V_{DD}$=0.6V.

Another factor that affects the sense margin and the search time is the bias voltage at evaluation transistor gate. 
To implement the functionality of bit-by-bit tunable threshold approximate matching, we determine appropriate evaluation voltages $\textit{V}_\textit{eval}$ to distinguish different mismatch thresholds, taking the threshold ranging 0-6 bits as an example.
This involves adjusting the gate voltage of the evaluation transistor to differentiate between 0-bit and 1-bit mismatch (Th-0), 1-bit and 2-bit mismatch (Th-1), and so forth. 
%adjacent numbers of mismatched bits can cause the $ML$ voltage to exhibit either a high level (maintained at 1V) or a low level (dropped to 0V) within the same time window.
%Once this time window is established, the array's capability for approximate matching can be verified by controlling the magnitude of the evaluation voltage.%To verify the proper functioning of the approximate match capability, it is essential to determine the appropriate evaluation voltage $V_{eval}$ to distinguish adjacent numbers of mismatched bits within the range of 0-6 bits. This involves adjusting the gate voltage of the evaluation transistor to differentiate between 0-bit mismatch and 1-bit mismatch, 1-bit mismatch and 2-bit mismatch, and so forth. The method of differentiation entails ensuring that within the same time window, the adjacent number of mismatched bits causes the voltage on the ML to exhibit either a high level (maintained at 1V) or a low level (dropped to 0V). Once the time window for distinguishing adjacent numbers of mismatched bits is established, we can verify that the array can achieve approximate match functionality by controlling the threshold voltage. 
Increasing the number of mismatch bits and evaluation transistor gate voltage $\textit{V}_\textit{eval}$ lead to faster $\textit{SA}_\textit{out}$ voltage decrease. 
Hence, with increasing mismatch threshold, we decrease $\textit{V}_\textit{eval}$ to maintain consistent sense time window across different mismatch thresholds.
%Based on this, we conducted experiments, obtaining different $V_{eval}$ values for matching thresholds ranging from 0 to 5 bits. 
The evaluation voltages are therefore experimentally examined and configured as summarized in \autoref{tab:Threshold-V} to ensure that the sense time for distinguishing different mismatch thresholds falls within the same time window.
Different evaluation voltages correspond to different mismatch thresholds. 
%\autoref{tab:Threshold-V} presents the evaluation transistor gate voltage for differentiating adjacent mismatched bit numbers, corresponding to different matching thresholds. 
This evaluation voltage configuration lays the foundation for subsequent performance and latency analysis.

%Different evaluation voltages correspond to different matching thresholds. As the number of mismatched bits increases and the gate voltage of the evaluation transistor $V_{eval}$ increases, both contribute to a faster decrease in the output voltage $SA_{out}$. Consequently, as the matching threshold increases, we gradually decrease the value of $V_{eval}$ to ensure a consistent sensing time window across different match thresholds. Based on this, we have conducted experiments and obtained different $V_{eval}$ corresponding to matching thresholds ranging from 0 to 5 bits. \autoref{tab:Threshold-V} presents the evaluation transistor gate voltage corresponding to distinguishing adjacent numbers of mismatched bits, i.e., the evaluation voltage corresponding to different matching thresholds. These findings will serve as the foundation for our subsequent performance and latency analysis.

\begin{table}[!t]
    \centering
    \caption{$\textit{V}_\textit{eval}$ of different mismatch threshold}
    \begin{adjustbox}{center}
    \resizebox{1\columnwidth}{!}{
\begin{tabular}{|c|c|c|c|c|c|c|}
              \hline
            \makecell{Mismatch\\Threshold(bit)} & 0 & 1 & 2 & 3 & 4 & 5   \\
              \hline
              $\textit{V}_\textit{eval}$(V) & 1 & 0.75 & 0.63 & 0.52 & 0.43 & 0.37  \\ 
              \hline
\end{tabular}

}
\label{tab:Threshold-V}
\end{adjustbox}
\end{table}
\begin{figure}
    \centering
    \includegraphics[width=\linewidth]{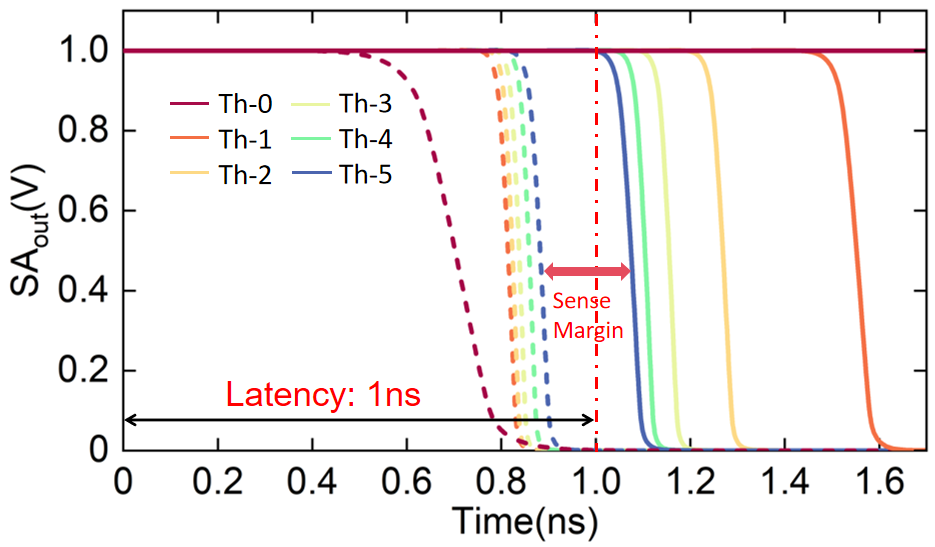}
    \caption{Transient waveforms of \textit{ML} under different mismatch thresholds. Solid and Dashed lines represent the match and mismatch cases corresponding to a certain mismatch threshold, respectively.}
\label{fig:threshold}
%\vspace{-0.4cm}
\end{figure}

The \textit{ML} transient waveforms corresponding to different mismatch thresholds in \autoref{fig:threshold} validate the bit-by-bit tunable threshold matching function.
%variations and the sensing time window during approximate matching under corresponding evaluation voltages and matching thresholds. 
Solid lines show the \textit{ML} voltage waveforms when the number of mismatched bits equals to the pre-defined mismatch threshold, while dashed lines show the \textit{ML} voltages when the number of mismatched bits  exceeds the pre-defined threshold. 
The sense margin of mismatch thresholds decreases as the threshold increases. 
%Notably, at Th-5, common window exists between 5-bits and 6-bits mismatch, with sensing time window endpoints serving as sensing margin. 
According to \autoref{fig:threshold}, the search latency for distinguishing adjacent mismatch threshold ranging from Th-0 to Th-5  is  1 ns.

\section{Evaluation}
\label{sec:eval}
In this section, we first evaluate the energy and performance of the proposed TAP-CAM design. We then benchmark the proposed TAP-CAM array in the context of K-nearest neighbor search tasks as tunable approximate matching engine.

\subsection{Evaluation Setup}

\begin{figure}
    \centering
    \includegraphics[width=\linewidth]{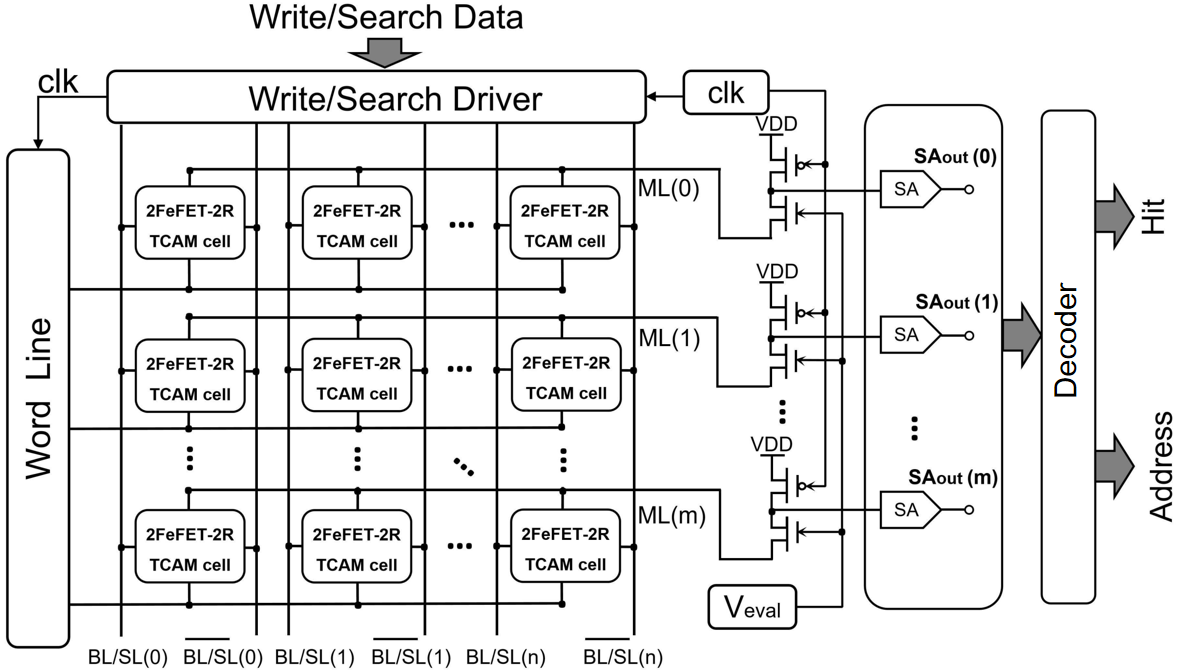}
    \caption{Schematic of m$\times$n TAP-CAM array.}
\label{fig:mxn}
\end{figure}

For the energy and performance evaluations, we conduct our experiments on
%original 1x64 array of 2FeFET-2R TCAM cells into 
a TAP-CAM array with m rows and n columns, as shown in \autoref{fig:mxn}. 
%Expanded array accommodates m words, each of length n.
The cells within the same row share the \textit{ML} and \textit{ScL}, and %the array possesses identical functionalities. 
the cells within the same column share \textit{SLs}, enabling parallel search operations. 
%while cells in the same row share $ML$ voltages, controlling $ML$ voltage variations simultaneously. 
Write/Search buffer drive stored/search vectors into \textit{SLs} for search operations, consistent with \autoref{tab:write opetarion}. 
During the search, all rows compare the same input query with stored entries.
If a mismatch occurs, \textit{ML} discharges. 
If \textit{ML} voltage drops below the sense amplifier threshold within the pre-defined sense time window, the corresponding SA output transitions to 0, recognized by the decoder as mismatch. Conversely, if a match occurs, the address of the stored entry matching the search query is output.

%Compared to a single-row 2FeFET-2R TCAM, the expanded m×n array has larger capacity and efficiently executes matching operations during search. By sharing $SLs$ and $ML$ voltage, the array achieves highly parallel search operations, thereby improving overall performance and efficiency.
The proposed 2FeFET-2R TAP-CAM array is evaluated using SPECTRE. The FeFETs are simulated based on the Preisach FeFET model \cite{transfer-characteristics}.
All MOSFETs are modeled using the 45nm PTM model and the 27°C TT process corner \cite{evaluation2}. The wordlength is set to 64 cells.
%, and the write voltage is ±4V. 

\subsection{Robustness Validation}

\begin{figure}
    \centering
    \includegraphics[width=\linewidth]{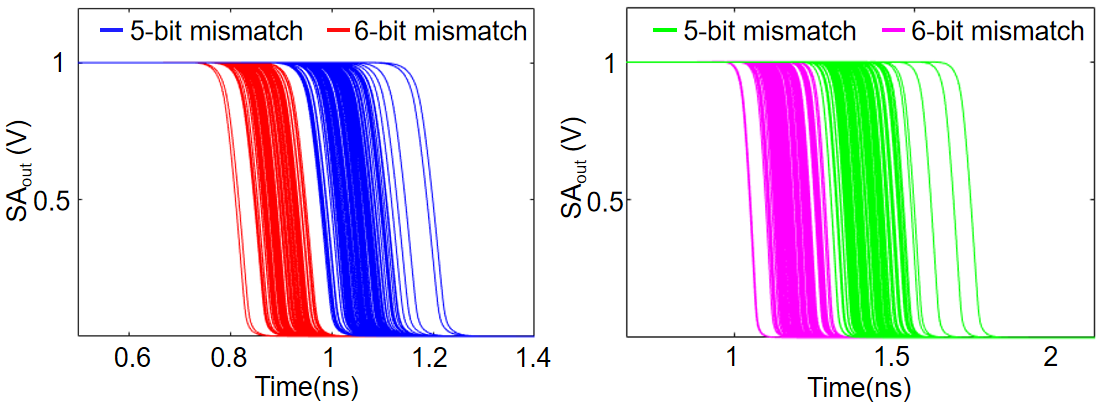}
    \caption{100 Monte Carlo simulations considering device-to-device variations: \textbf{(a)} The output waveforms under \textit{VDD} = 0.6V; \textbf{(b)} The output waveforms under \textit{VDD} = 1V.}
\label{fig:MC}
%\vspace{-0.2cm}
\end{figure}

The robustness of the proposed TAP-CAM design under varying operating conditions is examined, specifically with \textit{VDD} = 0.6V and \textit{VDD} = 1V, respectively. 
%To control the experimental variability of the FeFETs,
The FeFETs are assumed to feature the stored low/high $\textit{V}_\textit{TH}$ threshold voltage states with a deviation $\sigma$ = 54mV, and 8$\%$   series resistor variability is considered \cite{area}.
100 Monte Carlo simulations have been conducted  to distinguish between 5-bits and 6-bits mismatches when the mismatch threshold is set to 5 bits (Th-5). 
\autoref{fig:V&N} consistently reveals that the time windows across the 100 runs can be identified.
This observation suggests that the proposed design effectively distinguishes between the adjacent numbers of mismatched bits by employing the evaluation transistor. 
%Furthermore, the design maintains consistent search performance under different experimental conditions. 
Based on these results, it can be inferred that the proposed TAP-CAM design  demonstrates the robustness, as it reliably achieves approximate threshold matching functionality given the variations in operating voltage and device variations.

\subsection{CAM Array Evaluation}
%Here we conduct the evaluations of our 
\begin{figure}
    \centering
    \includegraphics[width=\linewidth]{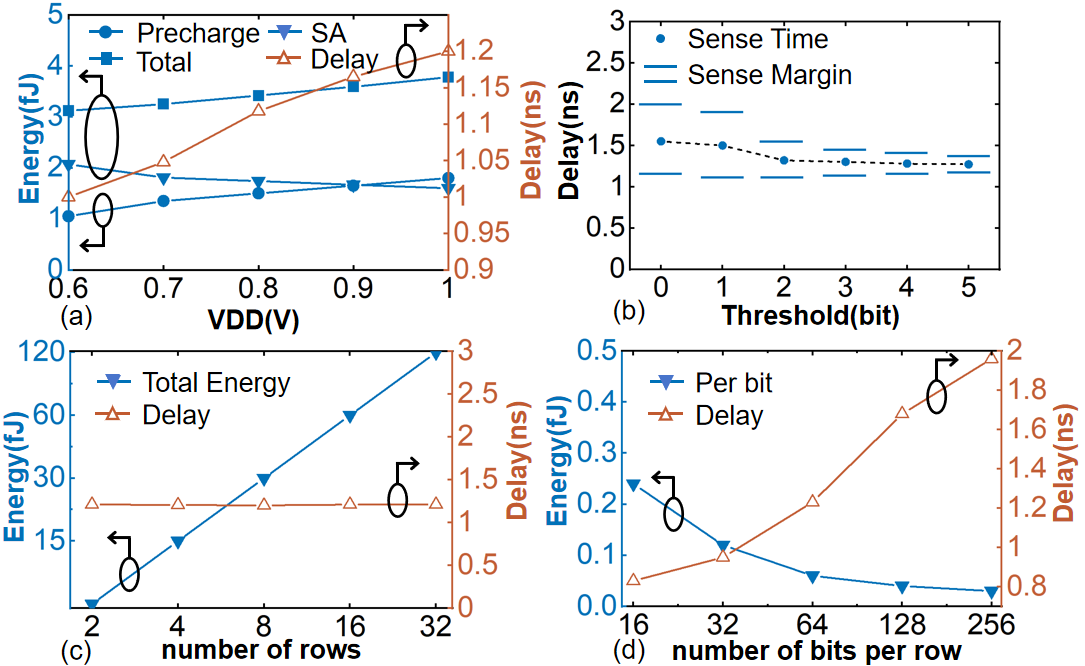}
    \caption{Energy and latency of the proposed 2FeFET-2R TAP-CAM array
with varying \textbf{(a)} \textit{VDD}; \textbf{(b)} mismatch thresholds; \textbf{(c)} number of rows and \textbf{(d)} number of bits per row.}
\label{fig:V&N}
%\vspace{-0.4cm}
\end{figure}

The search energy consumption of the proposed array mainly originates from precharging the \textit{ML} and SA energy consumption.
Precharging the \textit{ML}, primarily done by $\textit{T}_\text{1}$, depends heavily on \textit{VDD} and the associated \textit{ML} parasitic capacitance.
\autoref{fig:V&N}(a) demonstrates the impact of scaling \textit{VDD} on the search energy consumption and latency. 
As \textit{VDD} scales up, the precharging energy increases, leading to overall higher search energy consumption. At the same time, the amplitude of \textit{ML} dropping from high to low level when mismatch occurs increases, thereby increasing the search delay.  
\autoref{fig:V&N}(b) shows the sense time and sense margin for different mismatch thresholds at \textit{VDD} = 1V. The sense margin is the narrowest at the 5-bit mismatch threshold (Th-5), thus is selected as the sense margin for the SA sense time. 
\autoref{fig:V&N}(c) demonstrates how search energy and latency change with varying row numbers. Increased rows allow parallel search operations, linearly increasing the energy consumption with negligible latency change. 
Finally, \autoref{fig:V&N}(d) examines the wordlength's effect on the search latency and energy consumption per bit. 
Longer wordlengths associate more parasitic capacitance on the \textit{ML}, slowing down the discharge speed and thus increasing the search latency. The increase in capacitance 
leads to a rise in precharge energy per word. But increasing wordlength has minimal impact on the energy consumption of SA, so the search energy per bit decreases. The increasing latency and decreasing energy consumption per bit show trade-offs in the CAM array design optimization.

%The energy consumption in this process predominantly originates from two sources: the precharging of the $ML$ and the energy consumption by the SA. The precharging of the $ML$, primarily accomplished by $T_1$, is heavily contingent upon the magnitude of $V_{DD}$ and the parasitic capacitance. 

%\autoref{fig:V&N}(a) illustrates the impact of increasing $V_{DD}$ on energy consumption and latency. As $V_{DD}$ increases, precharging energy also increases, leading to overall higher energy consumption and latency. 
%\autoref{fig:V&N}(b) presents the sensing time and sensing margin for different matching thresholds at $V_{DD}$=1V. Notably, the sensing margin is narrowest at the 5-bit matching threshold, which determines the final sensing margin. \autoref{fig:V&N}(c) demonstrates how energy and latency vary with different numbers of rows. With increased row numbers, multiple rows can operate simultaneously, linearly increasing power consumption without affecting latency.
%Finally, \autoref{fig:V&N}(d) examines the effect of word length on latency and energy consumption per bit. Longer word lengths result in higher parasitic capacitance on the $ML$, slowing discharge speed and increasing latency. Conversely, longer word lengths lead to lower energy consumption per bit, highlighting the trade-offs in CAM array design optimization.

\begin{table}[!t]
    \centering
    \caption{Metric Comparison Summary of CAM Designs}
    \label{tab:example}
    \begin{adjustbox}{center}
   \resizebox{1\columnwidth}{!}{
        \begin{tabular}{|c|c|c|c|c|c|}
          \hline \hline
          Reference & \cite{16TCMOS}, \cite{2FeFETa} & \cite{conventionalCAM} & \cite{MASC} & \cite{2FeFETa} & Our Work   \\ 
          \hline
          Technology & CMOS & CMOS & ReRAM & FeFET & FeFET  \\ 
          \hline
          Node(nm) & 45 & 65 & 45 & 45 & 45  \\ 
          \hline
          Transistors/cell & 16T & 10T & 2T-2R & 2FeFET & 2FeFET-2R\\ \hline
          Match Style & Exact & Threshold & Threshold & Threshold &   Threshold\\
          \hline
          Cell size($\mu m^{2}$) & 1.2 & 5.45 & 0.41 & 0.15 & 0.15${^*}$ \\
          \hline
          Search delay(ps) & 582 & 1000 & 1450 & 355 & 1200 \\
          \hline
          \makecell{Energy\\ (fJ/bit/search)} & \makecell{1.00\\16.95$\times$} & \makecell{0.76\\12.88$\times$} & \makecell{0.56\\9.49$\times$} & \makecell{0.4\\6.78$\times$} & \makecell{0.059\\1$\times$} \\ \hline
          \hline
        \end{tabular}
     %    \begin{tablenotes}
     %   \footnotesize
    %  \end{tablenotes}
        }
    \end{adjustbox}
             \flushleft *: Back-end-of-line resistor incurs no additional area overhead as reported in \cite{area}.
    \label{tab:compare}
\end{table}

\autoref{tab:compare} provides a comprehensive comparison of the proposed 2FeFET-2R TAP-CAM with other CAM designs, in terms of  device type, technology node, device count per cell, cell size, performance and normalized search energy. Cell size estimation is based on a 2$\times$2 layout of the 2FeFET-2R TAP-CAM array.
%\autoref{tab:compare} presents a comprehensive comparison of the 2FeFET-2R CAM with other CAM designs, including the utilized technology, node, device count per cell, cell size, search delay, and search energy per bit per search.  The cell size estimation is based on a 2x2 layout of the 2FeFET-2R CAM array. 
%Our design leverages innovative FeFET technology, enabling approximate matching through different thresholds, achieving significant breakthroughs in circuit functionality and energy consumption. 
Compared to the conventional  CMOS  CAM designs, our proposed  2FeFET-2R TAP-CAM design offers a much smaller cell size. 
The  comparisons highlight the significant advantages of the proposed 2FeFET-2R TAP-CAM design over other CAM designs in terms of energy consumption per bit per search.
The energy efficiency of 2FeFET-2R TAP-CAM is notably superior, being 16.95$\times$, 12.88$\times$, 9.49$\times$, and 6.78$\times$ more efficient compared to 16T TCAM, 10T CAM, 2T-2R TCAM, and 2FeFET TCAM, respectively. 
While some existing designs  achieve approximate search functionality, their energy consumption remains substantially higher than that of 2FeFET-2R structure. 
%Utilizing ferroelectric transistor technology, 2FeFET-2R design represents promising direction for CAM design innovation, providing practical solution for enhancing search efficiency, reducing energy consumption, and optimizing area costs.
Although our design incurs relatively high search delay, considering the search latency and energy  trade-offs and the substantial energy advantages of our proposed design, increased delay is deemed acceptable.

These findings validate the remarkable energy efficiency of 2FeFET-2R TAP-CAM  array, emphasizing its immense potential for data-intensive search applications. This suggests that 2FeFET-2R TAP-CAM architecture is well-positioned to address the evolving needs of modern computing environments, particularly those requiring efficient and high-performance solutions for processing large volumes of data in search-intensive applications.

%Our design leverages innovative FeFET technology, enabling approximate matching through different thresholds, and has achieved significant breakthroughs in both circuit functionality and energy consumption. Compared to CMOS technology, the 2FeFET-2R structure offers a smaller cell size, thus reducing area costs. Although our design incurs a relatively high search delay, considering the trade-off between latency and energy consumption, as well as the substantial energy advantages of the 2FeFET-2R structure, the increased search delay is deemed acceptable. The comparison provided highlights the significant advantages of the 2FeFET-2R CAM structure over several other CAM designs in terms of energy consumption per bit per search. The energy efficiency of the 2FeFET-2R CAM is notably superior, being 16.95$\times$, 12.88$\times$, 8.81$\times$, and 6.78$\times$ more efficient compared to 16T CAM, 10T CAM, 4T-2FeFET CAM, and 2FeFET CAM, respectively. Despite the capability of some alternative designs to achieve approximate search functionality, their energy consumption remains substantially higher than that of the 2FeFET-2R structure. These findings validate the remarkable energy efficiency of the 2FeFET-2R CAM circuit array, emphasizing its immense potential for data-intensive search applications. Utilizing ferroelectric transistor technology, the 2FeFET-2R design represents a promising direction for CAM design innovation. It presents a practical solution for enhancing search efficiency, reducing energy consumption, and optimizing area costs.

\subsection{Case Study: K-Nearest Neighbor Search}
\begin{figure*}
    \centering
    \includegraphics[width=\linewidth]{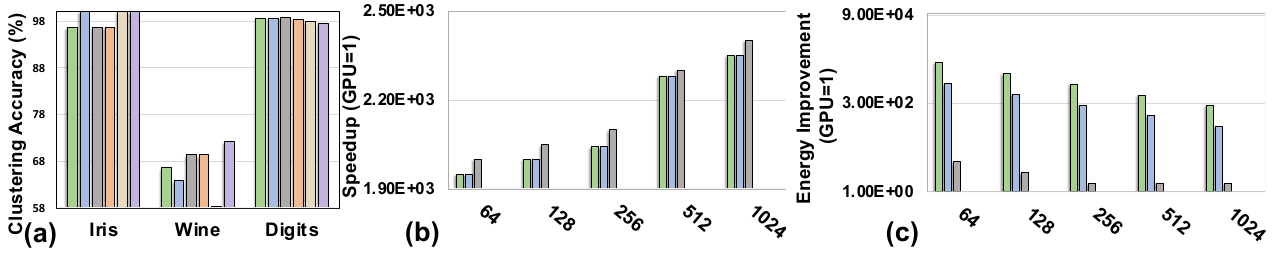}
    \caption{\textbf{(a)} KNN clustering accuracy under different \design thresholds, ranging from Th-1 to Th-6 (left to right); \textbf{(b)} Computational speedup and \textbf{(c)} energy efficiency improvement of \design with varying wordlengths compared to a GPU implementation. Datasets from left to right are Iris, Wine and Digits. }
    \label{fig:benchmark}
\end{figure*}
To demonstrate the efficiency of the proposed  design, we benchmark the proposed 2FeFET-2R TAP-CAM array in the context of  K-nearest neighbor (KNN) search framework. 
KNN, a fundamental algorithm in machine learning, embodies a non-parametric supervised model, particularly effective when $\textit{K = }1$, representing the nearest neighbor (NN) classification. 
This algorithm finds widespread use across various fields, including HDC  \cite{liu2022cosime, shou2023see}, reinforcement learning \cite{li2022associative}, and bioinformatics \cite{laguna2020seed}, etc.

At the core of the KNN approach lies the calculation of distances between the query instance, denoted as $x$, and the stored vectors, denoted as $y_i$, within the CAM array.
This process utilizes a distance function, typically denoted as $d(x, y_i)$, which quantifies the dissimilarity or similarity between the data points. 
When $\textit{K = }1$, i.e. NN classification, the class label attributed to the query instance $x$ corresponds to the category of the nearest stored vector $y_i$, identified by the smallest distance metric. This intuitive method allows for straightforward classification based on proximity, making it particularly suitable for scenarios with intricate decision boundaries or complex dataset patterns.
Conversely, when \textit{K} exceeds 1 instead of relying on the nearest neighbor, the algorithm considers the k closest neighbors of the query instance $x$. The class label assigned to $x$ is determined by a majority voting mechanism, where the most frequent class label among the k nearest neighbors prevails. 
This adaptive approach enables KNN to capture more nuanced relationships within the dataset, thereby enhancing its predictive capability and robustness in various applications.

%In the case of NN classification ($K=1$), the class label attributed to the query instance $x$ corresponds to the category of the nearest stored vector $y_i$, identified by the smallest distance metric. This intuitive method allows for straightforward classification based on proximity, making it particularly suitable for scenarios with intricate decision boundaries or complex dataset patterns.

%Expanding on this foundation, the KNN algorithm handles cases where K exceeds 1. In such situations, instead of relying on the nearest neighbor, the algorithm considers the K closest neighbors of the query instance $x$. The class label assigned to $x$ is determined by a majority voting mechanism, where the most frequent class label among the K nearest neighbors prevails. This adaptive approach enables KNN to capture more nuanced relationships within the dataset, thereby enhancing its predictive capability and robustness in various applications.

%In benchmarking our proposed 2FeFET-2R TAP-CAM architecture within the KNN framework, our goal is to showcase its versatility, efficiency, and applicability across various machine learning tasks. Through thorough evaluation and comparison with existing methodologies, we aim to highlight the potential of our design to advance CAM technology and contribute to machine learning research and development.

In benchmarking our proposed 2FeFET-2R TAP-CAM, for a given a function $d(x,y_i)$, which measures the distance between the query $x$ and the i-th stored vector $y_i$ in the CAM array, NN assigns the class label with the smallest distance value to $x$. Similarly, in KNN, given a query $x$, it assigns the most common class label of $x$'s k nearest neighbors to $x$ \cite{jiang2007survey}, as illustrated in \autoref{eq:knn}.
\begin{equation}
\label{eq:knn}
    c(x) = argmax\ \sum^k_{i=1} \delta(c,c(y_i))
\end{equation}
where $c(x)$ represents the class label of the query $x$, while $c(y_i)$ represents that of $y_i$. $y_i$ with $i$ ranges from 1 to k represent the k nearest neighbors. We have $\delta(c,c(y_i))=1$ when the query's label $c$ equals the label of $y_i$, otherwise $\delta(c,c(y_i))=0$.

\begin{table}[!t]
% \vspace{-3mm}
\centering
\caption{Datasets ($n$: total instances, $f$: features, $K$: number of classes)}
% \vspace{-3mm}
\label{tab:benchmark}
\resizebox{1\columnwidth}{!}{
\begin{tabular}{c|cccc}
\toprule
\textbf{Dataset}& $n$ & $f$ &$K$ & \textbf{Description}             \\ \midrule
\textbf{Iris} & 150               & 4          & 3                & Species of Iris \cite{dataset}\\ 
\textbf{Wine} & 178              & 13&3         &   Chemical analysis of wines \cite{dataset}       \\
\textbf{Digits} &  5620            &  64        &  10     & Hand-written digits \cite{dataset}          \\
\bottomrule
\end{tabular}
}
% \vspace{-1ex}
\end{table}
To comprehensively evaluate the effectiveness and performance of the proposed TAP-CAM architecture, KNN clustering analysis is conducted under the three most frequently referenced datasets in the UCI Machine Learning Repository, as shown in \autoref{tab:benchmark}. The datasets include Iris, Wine, and Digits, representing a wide range of data types and complexities. In order to achieve a robust evaluation, we have partitioned these datasets into training sets and test sets at an 8:2 ratio to ensure accurate testing and comparison of TAP-CAM model's performance.

\autoref{fig:benchmark}(a) illustrates the effectiveness of the proposed \design architecture across different datasets. 
Among Iris, Wine, and Digits, the \textit{Wine} dataset exhibits the highest susceptibility to hardware device-level variations. This observation emphasizes the importance of robustness in hardware designs, particularly in applications where environmental factors introduce variability. Additionally, we have examined the accuracy performance of KNN search under different TAP-CAM thresholds. 
Interestingly, the results indicate that identifying the nearest neighbor may not always yield the optimal solution. 
For instance, the Iris, Wine, and Digits datasets achieve their respective maximum clustering accuracies at $\textit{K = }2$, $\textit{K = }6$, and $\textit{K = }3$, respectively. 
With the proposed tunable approximate matching scheme, an average 3.06 \% accuracy improvement is observed compared to existing exact-match CAM methods.

Power consumption is obtained via the \textit{Nvidia-smi} toolkit, with the study conducted on \textit{Nvidia 2080ti GPU}, and the \design operations are analyzed via the \textit{Pytorch profiler}. Assuming 256 \design rows, feasible in current manufacturing technology, the KNN clustering benchmark considers different \design wordlengths at the algorithmic level. Idling power is excluded from the results. \autoref{fig:benchmark}(b) illustrates that \design exhibits at least $1.95\times 10^3$ speedup compared to GPU implementation. 
In addition, the energy consumption in \design grows linearly with the number of cells per row, whereas GPU implementations show little increase with dimensionality increment.
Consequently, as dimensionality increases, energy efficiency improvement decreases as demonstrated in \autoref{fig:benchmark}(c). 
For the \textit{Digits} dataset,
\design energy increases with the large number of instances and features,
resulting in an average improvement of $3.15\times$ compared to GPU implementations.

These results illustrate the effectiveness of the proposed TAP-CAM architecture across multiple datasets and scenarios, confirming its feasibility and superiority in practical applications. Through evaluation and comparison with existing methodologies, we highlight the potential of our design to advance CAM technology and contribute to machine learning research and development.
% \vspace{-2ex}

\section{Conclusion}
\label{sec:conclusion}

In this paper, we introduce TAP-CAM, a compact and energy-efficient TCAM design capable of threshold approximate matching. 
We propose a novel 2FeFET-2R TCAM design which employs an evaluation transistor  to adjust the ML discharge rate and measure the Hamming distance between the input query and the stored entries. 
Through gate bias voltage configuration, TAP-CAM achieves bit-by-bit tunable HD threshold matching functionality that is a crucial operation in many data-intensive applications. Evaluation results and application benchmarking suggest that our proposed 2FeFET-2R TAP-CAM array surpasses other advanced CAM technology in both energy efficiency and performance.

\section*{Acknowledgements}
This work was supported in part by  NSFC (62104213, 92164203) and SGC Cooperation Project (Grant No. M-0612). 

\bibliographystyle{IEEEtran}
\bibliography{bib}
\end{document}